\documentclass[article,twocolumn,author-numerical,showpacs]{revtex4}

\usepackage{amsmath,amssymb}
\usepackage{graphicx}
\usepackage{dcolumn}
\usepackage{bm}
 \usepackage{epsfig,lipsum}
 \usepackage{color,soul}

\begin{document}
\title{Mass transport perspective on an accelerated exclusion process: \\
Analysis of augmented current and unit-velocity phases }
\author{Jiajia Dong$^{1}$, Stefan Klumpp$^{2}$, and R.K.P. Zia$^{3}$}
\address{$^{1}$ Department of Physics and Astronomy, Bucknell University, Lewisburg, PA 17837\\
$^{2}$ Max Planck Institute of Colloids and Interfaces, 14424 Potsdam, Germany \\
$^{3}$ Physics Department, Virginia Polytechnic Institute and State University, Blacksburg, VA, 24061 and \\
Department of Physics and Astronomy, Iowa State University, Ames, Iowa
50011}
\date{\today}
\begin{abstract}

In an accelerated exclusion process (AEP), each particle can ``hop'' to its adjacent site if empty as well as ``kick'' the frontmost particle when joining a cluster of size $\ell \leq \ell_\text{max}$. With various choices of the interaction range, $\ell_\text{max}$, we find that the steady state of AEP can be found in a homogeneous phase with augmented currents (AC) or a segregated phase with holes moving at unit velocity (UV). Here we present a detailed study on the emergence of the novel phases, from two perspectives: the AEP and a mass transport process (MTP). In the latter picture, the system in the UV phase is composed of a condensate in coexistence with a fluid, while the transition from AC to UV can be regarded as condensation. Using Monte Carlo simulations, exact results for special cases, and analytic methods in a mean field approach (within the MTP), we focus on steady state currents and cluster sizes. Excellent agreement between data and theory is found, providing an insightful picture for understanding this model system. 

\end{abstract}


\maketitle
\section{Introduction}

Unraveling rich behaviors emerging from simple ingredients in systems driven far from equilibrium is a continuous pursuit in theoretical physics. Non-trivial flux of physical quantities, or current, captures the macroscopic feature of the complex system and is often closely governed by the intrinsic dynamics. There are numerous examples where the steady state current depends sensitively on the constituents in the system through both long-range and short-range interactions such as queueing in traffic and pedestrians \cite{traffic}, transport of biomolecules \cite{cell,Epshtein03,Jin10,Lipowsky06} and minerals \cite{ion}. 

One of the venerable models, the totally asymmetric simple exclusion process (TASEP) not only provides many interesting mathematically exact results of non-equilibrium statistical mechanics \cite{Spitzer70, Derrida92, Derrida93, Schutz93,Schutz00}, it also brings insights to potentially important applications in, for example, protein synthesis \cite{MacDonald68,MacDonald69,Klumpp11,CMZ11} and regulating vehicular traffic \cite{Chowdhury00}. 
Closely related is the zero-range process (ZRP) in which particles hop between sites with rates determined by the origin site occupancy.  The mapping between ZRP and an ordinary TASEP proved illuminating in our later discussions. ZRP also finds its versatility in studying granular materials as well as phase separation in one-dimensional systems \cite{Evans05}.

Inspired by assisted hopping in transcription \cite{Epshtein03,Jin10}, we introduced a new variant of TASEP recently \cite{Dong12}, the ``accelerated exclusion process'' (AEP), to characterize interactions beyond nearest neighbors among particles. Violating detailed balance, AEP is non-equilibrium in nature and involves a few surprising phenomena. Let us briefly recapitulate the dynamic rules of TASEP before turning to the novel features of AEP: In each update attempt of TASEP, a particle in a discrete one-dimensional (1D) lattice of $L$ sites is chosen at random to hop into its neighboring site (provided that is vacant) with rate $\gamma$ (typically chosen as unity). Both periodic and open boundary conditions (particles enter and exit with rates $\alpha$ and $\beta$ respectively) have been studied extensively \cite{Spitzer70, Derrida92, Derrida93, Schutz93,Schutz00}. The non-equilibrium steady states of an ordinary TASEP are well-understood. For example, the current-density relationship is given by $J_\text{TASEP}=\rho(1-\rho)$, in the thermodynamic limit.

\begin{figure}[h]	
	\begin{center}
	\includegraphics[width=0.48\textwidth]{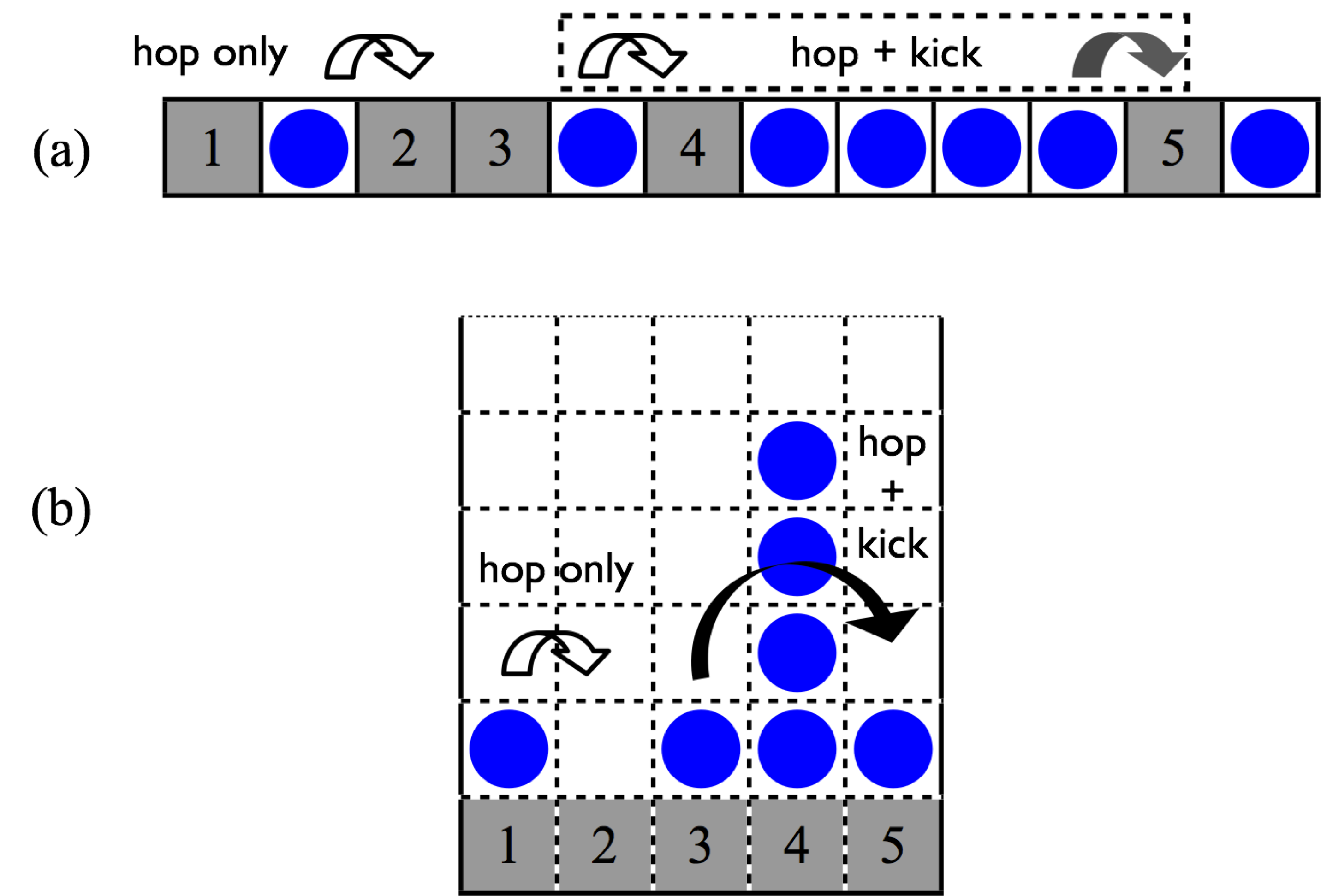}	
	\end{center} 
	\caption{(a) AEP with $\ell_\text{max}=4$. A particle can hop (hollow arrows) to its unoccupied right neighboring site \textit{and} kick (grey arrow) a second particle when joining a cluster of size $\ell\leq \ell_{\text{max}}$. (b) Mapping AEP in (a) to MTP. The ball in stack 3, if chosen, lands immediately in stack 5 (Òhop and kickÓ), while the one in stack 1 lands in stack 2 (ÒhopÓ).}
	\label{fig:aep}
	\end{figure}

In the AEP with periodic boundary condition, $N$ particles are placed in a ring of $L$ sites and follow the same rules as in TASEP. \emph{In addition}, when a particle hops to a cluster of particles of size $\ell\leq\ell_\text{max}$, it simultaneously ``kicks'' the frontmost particle of that cluster one site forward. There is no avalanche, as the ``kicked'' particle does not trigger another kick. When $\ell_\text{max}=0$, AEP reduces to an ordinary TASEP. A schematic of AEP is shown in Fig.~\ref{fig:aep}(a).

In this article, we focus our attention on the non-equilibrium steady states of this AEP and explore the interplay amongst the overall density $\rho\equiv N/L$, $\ell _\text{max}$ and current $J(\rho;\ell_\text{max})$. Due to ``kicking,'' a number of novel features arise in the AEP. 
Given that the contribution to the current can be either 1 (a hop) or 2 (a hop and a kick) \cite{Dong12}, the system is naturally expected to display an \textit{augmented current} (AC). 
In this state, the system is homogeneous and we may anticipate that $J_\text{AC}$ lies between $J_\text{TASEP}$ and $2\times J_\text{TASEP}$. It is intriguing that, at high densities ($\rho>1/2$), $J_\text{AC}$ can exceed $2\times J_\text{TASEP}$. 
Meanwhile, for low densities, the kicking action results in ``facilitated'' or ``cooperative motion'' \cite{AS00,Gabel10,GR11}, where the average velocity of the particles, $v$, can be increased by adding particles to the system. 
Clearly absent in the ordinary TASEP, this phenomenon can be characterized by $\partial_\rho v> 0$ or a positive curvature in $J$: $\partial_\rho ^2J> 0$ \cite{AS00,Gabel10,GR11}. 
Furthermore, for moderate values of $\ell_{\text{max}}$ and at high densities, the system exhibits an inhomogeneous state in which the particles ``condense'' into a macroscopic, ``solid'' cluster, in coexistence with a ``fluid'' of density $\sim 1/2$. 
Surprisingly, the (average) current is just $1-\rho = H/L$, where $H$ is the number of holes in the system. Thus, the fluid can be regarded as a loosely bound set of holes, moving together with unit velocity (UV). 
These two different states of the system will be referred to as the AC and the UV phases, respectively.
As the overall density $\rho$ is increased with fixed $\ell_\text{max}$, we observe a discontinuous jump in $J(\rho;\ell_\text{max})$, from a non-trivial $J_\text{AC}(\rho)$ to the simple $J_\text{UV}=1-\rho$. 
Our goal is to understand these remarkable phenomena.  

Although AEP is originally cast in the language of an exclusion process, both the intuitive picture and the analysis for predicting the aforementioned features turn out to be much easier when viewed in an equivalent representation, the \textit{mass transport process} (MTP).
In the next Section, we present a detailed description of the AEP-MTP mapping, which is a simple generalization of the TASEP-ZRP mapping. In this setting, the exact master equation can be easily written. Following a brief summary of simulation results in Section \ref{sec:sim}, we provide theoretical considerations for the properties of AC and UV phases in Sections \ref{sec:AC} and \ref{sec:UV}, respectively. In Section \ref{sec:trans}, we venture a phase diagram for this system. We conclude and provide an outlook for further quests in Section \ref{sec:sum}.

\vspace{0.5cm}

\section{Accelerated Exclusion as a Mass Transport Process\label{sec:MTP}}
Regarded as particles traversing a 1D ring, AEP allows particle to move \textit{only} when it is adjacent to a hole. The configuration of the system, $\mathcal{C}$, can be characterized by the set of site occupancies, $\{n_i\}, i=1,...,L$, with $n$ being 0 or 1. Alternatively and more conveniently, AEP can be formulated as a mass transport process (MTP) \cite{Kafri02} in which we regard the particles in front of each hole as ``balls'' stored in a ``stack.''  Each hole in AEP becomes a stack in MTP, labeled by $\alpha =1,...H$. The $\ell _\alpha $ balls in stack $\alpha$ correspond to
the cluster of particles between the $\alpha ^\text{th}$ and $(\alpha +1)^\text{th}$ hole. Each stack may be occupied by any number of (indistinguishable) balls: $\ell _\alpha =0,1,..., N$. An equivalent specification of $\mathcal{C}$ can thus be the set $\left\{ \ell _\alpha \right\} $ instead.

The mapping from AEP in Fig.~\ref{fig:aep}(a) to MTP is shown in Fig.~\ref{fig:aep}(b). The dynamic rules of AEP thus become: In each update attempt, a
random stack $\alpha$ is chosen. If it is empty ($\ell _\alpha =0$), another attempt is made. Otherwise, one of the balls in the chosen stack hops to the next stack 
{\it and} if $\ell _{\alpha +1}\in \left[ 1,\ell _{\max }\right] $, the ball takes
a second hop immediately, landing in stack $(\alpha +2)$. For instance, the ``hop and kick'' scenario in Fig.~\ref{fig:aep}(a) becomes a hop from stack 3 to 5 in Fig.~\ref{fig:aep}(b). When $\ell_\text{max}=0$, it returns to an ordinary TASEP.

Clearly, $N=\Sigma_\alpha \ell _\alpha $ is conserved. These rules are summarized in the
master equation for $P\left(\mathcal{C} ;t\right) $
in Eq.~(\ref{eq:ME}). One of the advantages of MTP representation is that $\left\{ \ell
_\alpha \right\} $ provides directly the cluster size distribution in the
AEP, a measure which we can use to characterize the AC and UV phases
quantitatively.

The only non-trivial aspect of this mapping is $J$, the overall particle
current. In each update attempt, a random site is chosen for AEP, while in MTP is a stack. Thus, the time scale differs by a factor of $H/L$. In
particular, if we compute the average number of particle movements in MTP, it must be
multiplied by $H/L=1-\rho $ when compared to $J$ in AEP.

We proceed to formulating the dynamic rules as a master equation. To facilitate this task, we define the characteristic functions 
\begin{eqnarray*}
\chi \left( \ell \right) &\equiv &1~~\text{if}~\ell\in \left[ 1,\ell _{\max }\right]
;~~0~~\text{otherwise}. \\
\overline{\chi}(\ell) &\equiv &1-\chi(\ell)\\
\end{eqnarray*}
Note that $\ell _{\max }$ is an implicit parameter in these functions. The master equation governing the evolution of $P\left(  \mathcal{C} ;t\right) $, namely the probability to find the system in configuration $\mathcal{C}=\left\{
\ell _a\right\} $ in $t$ attempts after some initial configuration, for $H>2$ is:

\begin{widetext}
\begin{equation}
\begin{split}
P\left( \mathcal{C}^{\prime }; t+1\right) =&\sum_{\mathcal{C}}\dfrac
1H\sum_{\alpha =1}^H P\left(\mathcal{C} ;t\right) \left[
\prod\limits_{\beta \neq \alpha ,\alpha +1,\alpha +2}\delta \left( \ell
_\beta ^{\prime },\ell _\beta \right) \right] \delta \left( \ell _\alpha
^{\prime },\ell _\alpha -1\right) \times \\
& \times \left\{ \delta \left( \ell _{\alpha +1}^{\prime },\ell _{\alpha
+1}\right) \delta \left( \ell _{\alpha +2}^{\prime },\ell _{\alpha
+2}+1\right) \chi \left( \ell _{\alpha +1}\right) +\delta \left( \ell
_{\alpha +1}^{\prime },\ell _{\alpha +1}+1\right) \delta \left( \ell
_{\alpha +2}^{\prime },\ell _{\alpha +2}\right) \overline{\chi}\left( \ell _{\alpha+1}\right) \right\}
\label{eq:ME}
\end{split}
\end{equation}
\end{widetext}

Here, $\delta $ is the Kronecker delta. For $H=1$, there is just one stack in MTP, leaving AEP trivial. For $H=2$, the problem is
easily solvable and hints at the significant role of $\ell _{\max }$. The
next case ($H=3$) is the first non-trivial one and the exact solution in the $%
L\rightarrow \infty, \ell _{\max }\rightarrow \infty $ limit provides
valuable insight into the UV phase. We show the details of $H=2$ and 3 in Appendices \ref{app:1} and \ref{app:2}.

The configuration space $\left\{ \ell _\alpha \right\} $
consists of the lattice points in an $(H-1)$-dimensional hyper-tetrahedron. The
easiest way to visualize this is its standard embedding in $H$-dimension,
i.e., a (linear) space joining the following $H$ points: $
(N,0,...,0),...,(0,...,0,N)$. For $H=3$ and 4, the configuration space
is just an equilateral triangle and the standard regular tetrahedron, respectively.

In general, this dynamics does not obey detailed balance, so that finding an
explicit stationary distribution, $P^{*}\left(\mathcal{C}\right) $, is not simple \cite{ZS07,Hill66}. To have some understanding of
the system behavior, we will exploit the approximation schemes presented
below, before which let us comment briefly on some general properties of our system.

Two extreme cases are noteworthy. One is $\ell _{\max }= 0$, which is
simply the ordinary TASEP. Dropping the $\chi $ term, Eq.(\ref{eq:ME}) reduces to 
\begin{equation}
\begin{split}
P_\text{TASEP}\left( \mathcal{C}^{\prime } ; t+1\right) =\sum_{\mathcal{C}}\frac{
1}{H}\sum_{\alpha =1}^{H}P_\text{TASEP}\left( \mathcal{C}; t\right)\times ~~~
\\
\times \delta \left( \ell _{\alpha }^{\prime },\ell _{\alpha }-1\right) \delta \left( \ell _{\alpha +1}^{\prime },\ell _{\alpha
+1}+1\right) \prod\limits_{\beta \neq \alpha ,\alpha +1}\delta \left( \ell
_{\beta }^{\prime },\ell _{\beta }\right) 
\end{split}
\end{equation}
Though this dynamics also violates detailed balance, it does satisfy the
``pairwise balance'' condition \cite{Spitzer70}, so
that $P_\text{TASEP}^{\ast }\left( \mathcal{C} \right)
\propto 1$. The opposite extreme is $\ell _{\max }\rightarrow
\infty $, or simply $\ell _{\max }>L$ or $N$. $\overline{\chi}\left( \ell
\right) $ reduces to $\delta \left( \ell \right) $ so that $\left\{
...\right\}$ in Eq.~(\ref{eq:ME}) becomes:
\begin{equation}
\begin{split}
\{ &\delta \left( \ell_{\alpha +1}^{\prime},\ell _{\alpha +1}\right) ~~~
\delta \left( \ell _{\alpha +2}^{\prime },\ell _{\alpha +2}+1\right) 
\overline{\delta }\left( \ell _{\alpha +1},0\right) +
\\
+& \delta \left( \ell_{\alpha +1}^{\prime },\ell _{\alpha +1}+1\right) \delta \left( \ell_{\alpha +2}^{\prime },\ell _{\alpha +2}\right) \delta \left( \ell _{\alpha+1},0\right) \}
\end{split} 
\end{equation}
where $\overline{\delta }\equiv 1-\delta $. This important limit exemplifies
the AC phase and will be examined more closely below.

Returning to the general case, one can compute the average of any quantity ${\cal Q}$
in the stationary state assuming $P^{\ast}$ is known:
\begin{equation*}
\left\langle {\cal Q}\right\rangle \equiv \sum {\cal Q}P^{\ast }
\end{equation*}
where the sum is taken over $\left\{ n_{i}\right\}$ or $\left\{ \ell
_{\alpha }\right\}$, whichever is more convenient. In particular, the
average particle current in steady state is given by:
\begin{equation}
\begin{split}
J\left( \rho ;\ell _{\max }\right)&=\left\langle n_{1}\left(
1-n_{2}\right) \left( 1+n_{3}\right) \right\rangle\\
&~~~~~   -\left\langle n_{1}\left(
1-n_{2}\right) \prod\limits_{m=0}^{\ell _{\max }}n_{3+m}\right\rangle \\
&=\left( 1-\rho \right) \left\langle \overline{\delta }\left( \ell
_{1},0\right) \left[ 1+\chi \left( \ell _{2}\right) \right] \right\rangle
 \label{AEPJ}
\end{split}
\end{equation}

Note that we have invoked translational invariance of $P^{\ast }$ in writing
these expressions. If the reader is concerned that this invariance may be
(spontaneously) broken, as in the UV phase, then $J=\sum_{\left\{ n\right\}
}\left\langle ...\right\rangle /L=$ $\sum_{\left\{ \ell \right\}
}\left\langle ...\right\rangle /H$ may be used instead. Also, the second line is the result of having one (two) hop(s) for a
particle-hole-hole(particle-hole-particle) triplet. In MTP, whenever stack 1
is chosen, a ball will hop, with an additional hop if the target stack has the
requisite number of balls.

\section{Simulation results\label{sec:sim}}
We exploit the random sequential updating scheme and simulate AEP with a typical lattice of $L=1000$. Initializing the system with $N$ particles on successive lattice sites, we make $N$ attempts to update in every Monte Carlo step (MCS) so that each particle has on average one chance to hop. The contribution to the current from both hops and kicks are instantly accounted for. The typical length of our simulations is $\tau=5\times 10^5$ MCS.  We start data collection after $2\times10^5$ MCS to ensure that the system has reached steady state and all measurements are averaged over $3\times 10^5$ MCS thereafter unless otherwise specified.

	\begin{figure}[h]	
	\begin{center}
	\includegraphics[height=6cm,width=8.5cm]{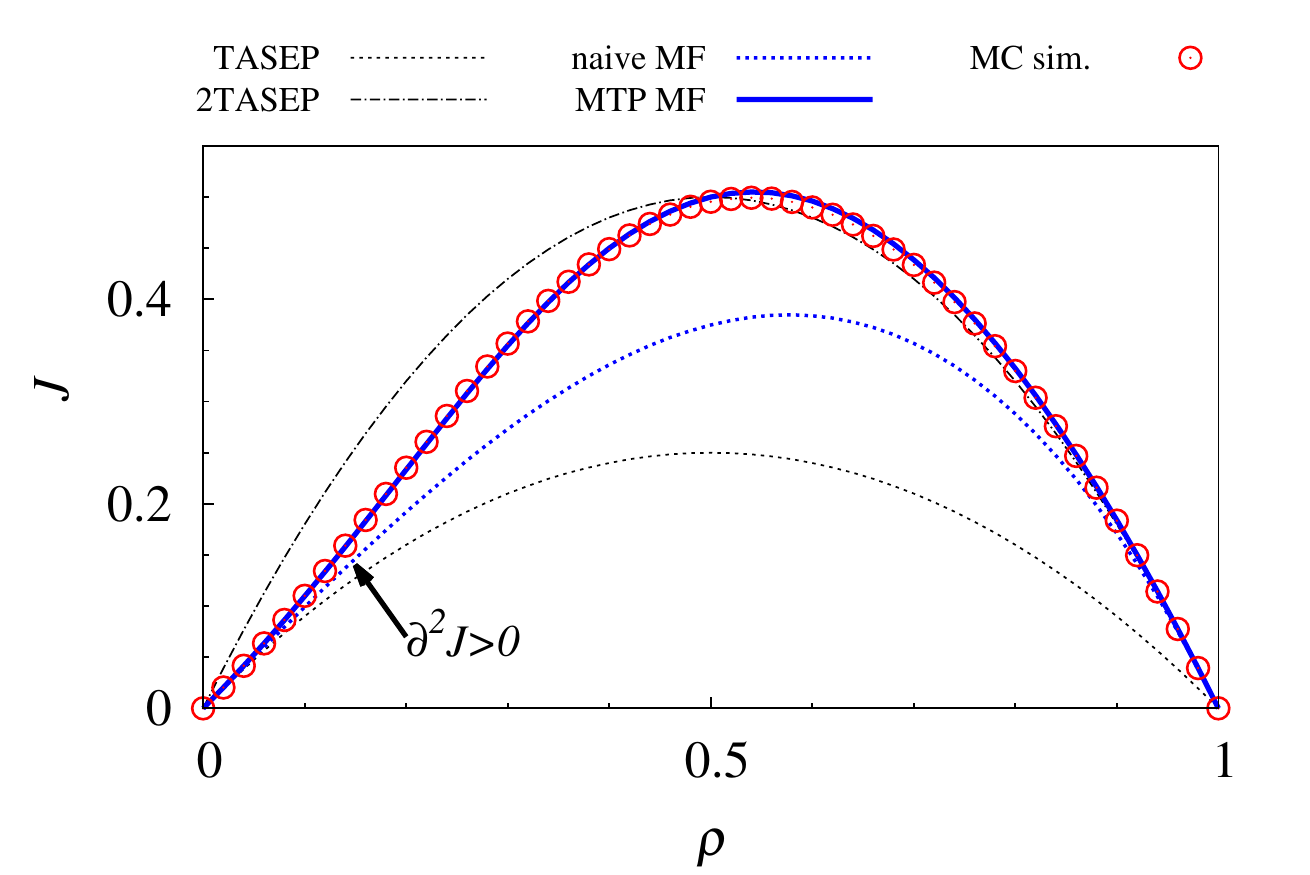}	
	\end{center} 
	\caption{(Color online) Comparison amongst $J_\text{AC}$ from Monte Carlo simulation, $J_\text{TASEP}$, $2J_\text{TASEP}$, a  na\"{\i}ve mean field approximation and an improved mean field approximation through MTP (MTP-MF). The result from MTP-MF in Eq.~(\ref{J-ACMFA}) (blue solid line) provides remarkable agreement with the simulation.}
	\label{fig:jac}
	\end{figure}
	\begin{figure}[h]	
	\begin{center}
	\includegraphics[height=6cm,width=8.5cm]{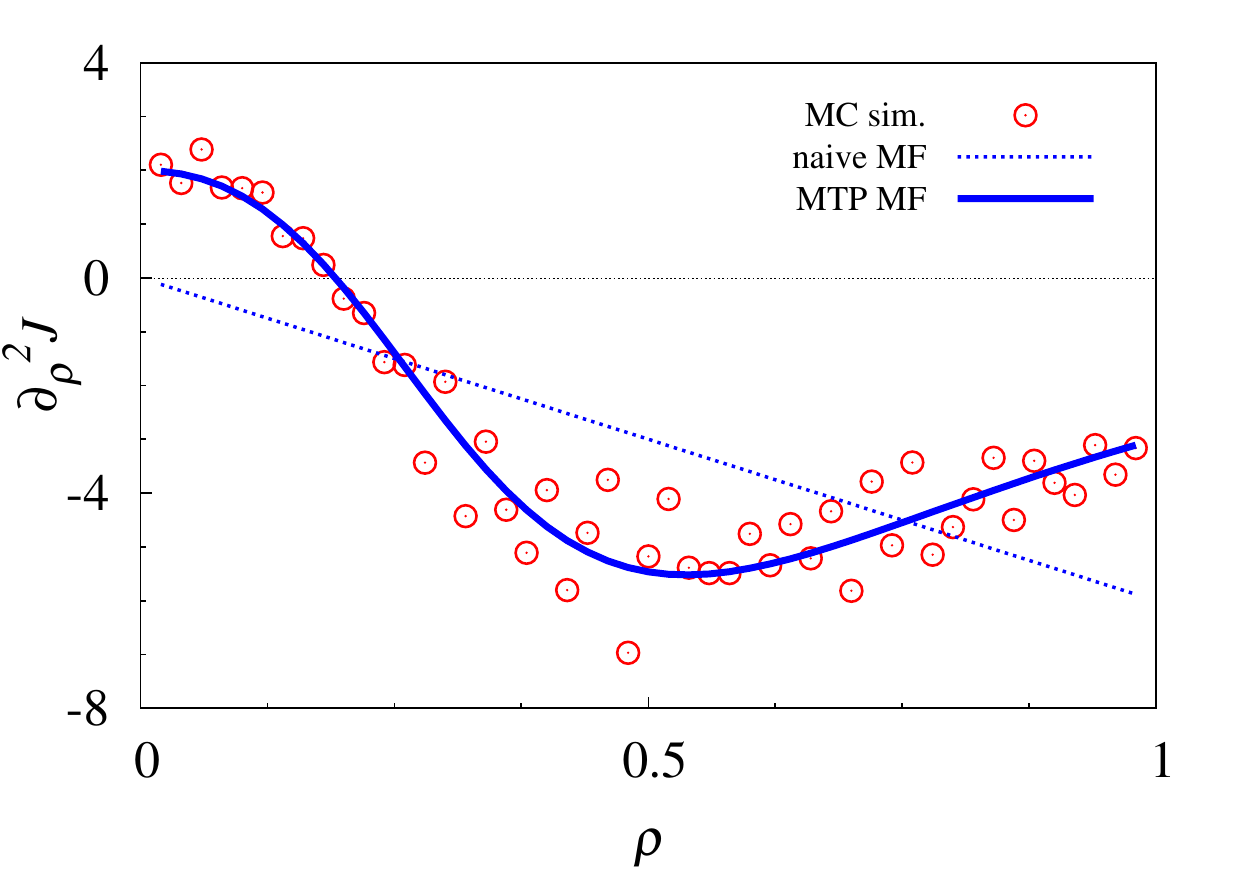}	
	\end{center} 
	\caption{(Color online) Curvature of $J(\rho)$ from simulation (circles),  a  na\"{\i}ve mean field approximation (blue dotted line) and MTP-MF (blue solid line).}
	\label{fig:curve}
	\end{figure}

First let us examine the AC phase. We note that not only is $J$ always larger than $J_\text{TASEP}$ except at $\rho=0$ and 1, it even surpasses $2J_\text{TASEP}$ when $\rho>1/2$, as shown in Fig.~\ref{fig:jac}.  The simulation (red circles) is for $\ell_\text{max}=1000$ and thus all hops lead to kicks. The system is always in the AC phase regardless of the density, which may lead one to believe that $J_\text{AC}$ should be close to $%
2J_\text{TASEP}=2\rho \left( 1-\rho \right) $, since a typical particle will hop
two sites instead of just one. This rough estimate is reasonably good for high densities, as Fig.~\ref{fig:jac} (black dot-dash line) indicates. Using the most na\"{\i}ve
mean field approximation, we can start from the exact expression Eq.~(\ref{AEPJ})
and let $\ell _{\max }\rightarrow \infty $. This leads us to $\rho
\left( 1-\rho ^{2}\right) $.  Surprisingly, this provides a much poorer
overall picture (blue dotted line in Figs.~\ref{fig:jac} and \ref{fig:curve}). 
Furthermore, there is some inflection at low density where the curvature of $J_\text{AC}(\rho)$ is positive, indicating an ``accelerated region.'' This is illustrated in Fig.~\ref{fig:curve}. 

Utilizing the MTP picture, we provide a mean field theory (MTP-MF) in Section \ref{sec:AC} to describe the AC phase, which yields exceptional agreement with simulation data, shown in solid blue line in Figs.~\ref{fig:jac} and \ref{fig:curve}.  This approach leads to a remarkably good description of all phenomena presented here. Moreover, it provides a viable explanation for
why the estimate $\rho \left( 1-\rho ^{2}\right) $ fails. The details are presented in Section \ref{sec:AC}.

\begin{figure}[h]	
	\begin{center}
	\includegraphics[height=6cm,width=8.5cm]{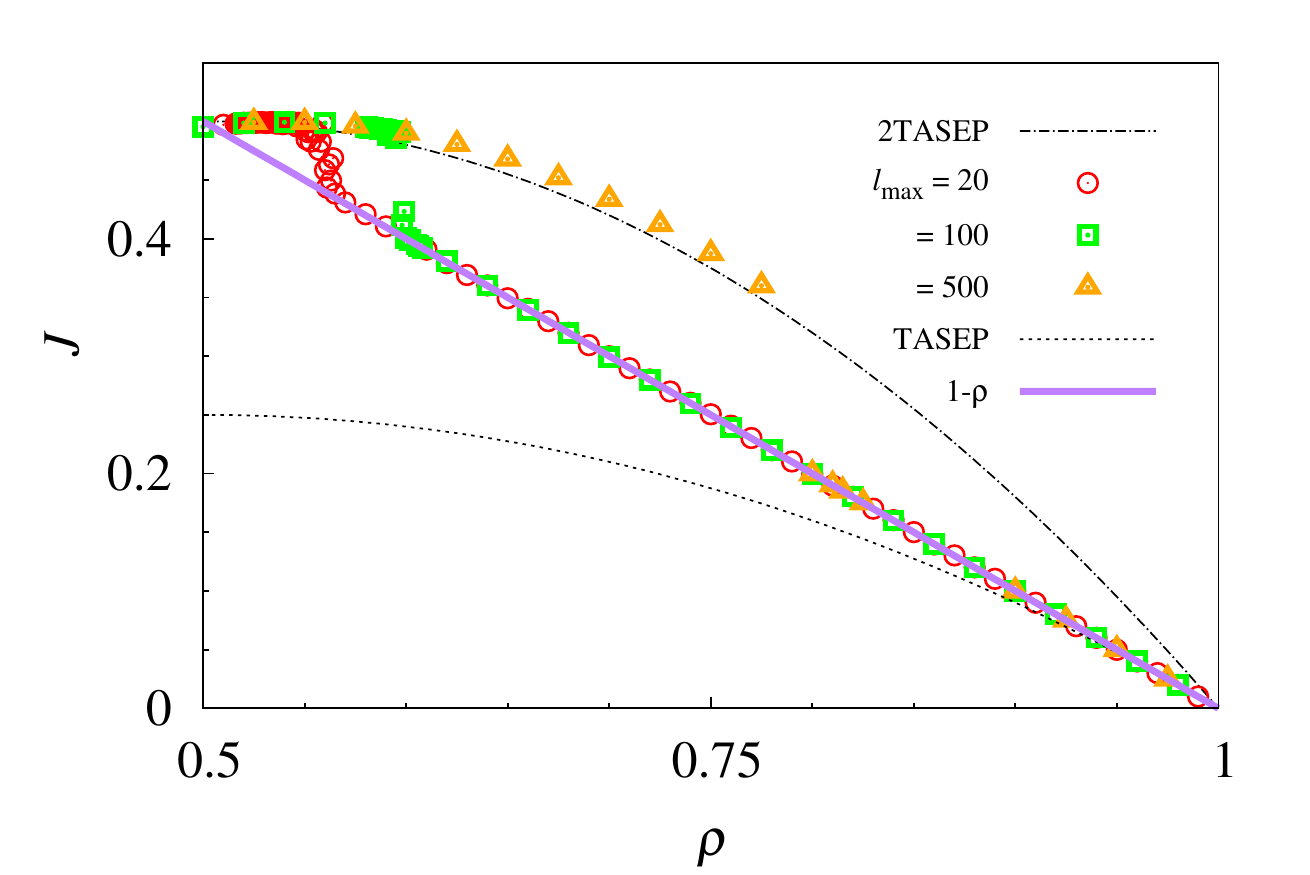}	
	\end{center} 
	\caption{(Color online) Transition from AC to UV for $\ell_\text{max} = 20$ (red circles), 100 (green squares) and 500 (yellow triangles). }
	\label{fig:juv}
	\end{figure}

The emergence of the UV phase depends on the appropriate choice of $\ell_\text{max}$. In all of our simulations where $L=1000$, there exists a UV phase for $10\lesssim\ell_\text{max}\lesssim950$. We show a few typical $\ell_\text{max}$'s in Fig.~\ref{fig:juv}. When the system enters UV, $J$ becomes independent of $\ell_\text{max}$ and is simply $(1-\rho)$, indicating the holes are moving at average speed 1. It is therefore both more effective and illuminating to switch to the reference frame of moving holes: Starting from a full system with one giant cluster of particles, i.e. $H=0$, as more and more holes are injected to the system, we find the system settles into a cluster of particles in one part and a half-filled region in the other part of the system. Exploiting both AEP and MTP frameworks, we explicate the emergence of such ``phase separation'' in Section \ref{sec:UV}.
\section{Augmented Current Phase: Mean Field Description\label{sec:AC}}

In this section, we focus on the stationary state of the AC branch, in which
the system is homogeneous and the density profile is just $\rho $. Now, if
we let $\ell _{\max }>L$, then the system be in AC for {\em all} $%
\rho $'s. In the MTP representation, the rule is especially straightforward: Every hop
into an occupied stack makes a further hop. Thus, the probability that a stack is occupied 
\[
f\equiv \left\langle \overline{\delta }\left( \ell _1,0\right) \right\rangle 
\]
will play a central role. Further, we find it useful to study the more
detailed distribution 
\[
P^{*}\left( \ell \right) \equiv \left\langle \delta \left( \ell ,\ell
_1\right) \right\rangle 
\]
i.e., the probability that a stack contains precisely $\ell $ particles. Of
course, $f=\sum_{\ell >0}P^{*}\left( \ell \right) $ implies 
\begin{equation}
1-f=P^{*}\left( 0\right) .  \label{1-f}
\end{equation}

\subsection{Steady state occupations}

Although it is not possible to derive an exact master equation for $P\left( \ell
;t\right) \equiv \sum \delta \left( \ell ,\ell _{1}\right) P\left(\mathcal{C} ;t\right) $ from Eq.~(\ref{eq:ME}), we will
exploit a mean field approach to find $P^{\ast }\left( \ell \right) $. To
maintain a steady state, a given stack must gain and lose a ball with equal
probability in any attempt. In other words, we can find $P^{\ast }$ by
balancing the average rates of gain and loss. 

Clearly, the probability that
an occupied stack ($\ell \geq 1$) loses a ball is $P^{\ast }\left( \ell
\right) $ (with the trivial factor of $1/H$ suppressed). Meanwhile, the stack
can gain one ball (i.e., $\ell -1\rightarrow \ell $) if both stacks upstream are
occupied. Injecting the mean field approximation, we estimate this condition
by $f^{2}$. If the chosen stack is empty, there is an {\em additional} way it can
gain, from its immediate upstream neighbor if occupied. Thus, the
balance equations read:
\begin{eqnarray*}
\left( f^{2}+f\right) P^{\ast }\left( 0\right) &=&P^{\ast }\left( 1\right)
\label{P01}\\
f^{2}P^{\ast }\left( \ell \right) &=&P^{\ast }\left( \ell +1\right) ;\,\ell
>0.\,\,
\end{eqnarray*}%
Strictly, we should account for the upper limit $\ell \leq N$. If we neglect
such finite size size effects, the normalization constraint is $%
\Sigma _{\ell =0}^{\infty }P^{\ast }\left( \ell \right) =1$, leading us to
an explicit expression for the stationary distribution: 
\begin{eqnarray}
P^{\ast }\left( \ell >0\right) &=&\left( 1+f\right) f^{2\ell-1}P^{\ast }\left(
0\right)  \label{P*ell} \\
P^{\ast }\left( 0\right) &=&1-f  \label{P*0}
\end{eqnarray}%
Note that Eq.~(\ref{P*0}) is entirely consistent with Eq.~(\ref{1-f}).

Next, let us relate $f$ to the density $\rho $. Computing $\left\langle \ell
\right\rangle $ is straightforward: $f/\left( 1-f^{2}\right) $. But this
average occupation must be $N/H=\rho /\left( 1-\rho \right) $, leading to%
\begin{equation}
f^{2}+\frac{H}{N}f=1  \label{f^2}
\end{equation}%
Instead of a cumbersome algebraic expression for $f\left( \rho \right) $, let
us define $\eta $ by 
\begin{equation}
\sinh \eta \equiv \frac{H}{2N}=\frac{1-\rho }{2\rho }  \label{eta}
\end{equation}%
and recognize $\left( 1-f^{2}\right) /f=f^{-1}-f$, so that \begin{equation}
f=e^{-\eta }  \label{f-eta}
\end{equation}%
or explicitly, $f=\exp \left\{ -\sinh ^{-1}\dfrac{1-\rho }{2\rho }\right\} $ \footnote{%
If we had used TASEP rules, we would find $\left\langle \ell \right\rangle
=f/\left( 1-f\right) $ and so, $f=\rho $.} 
. Not surprisingly, $f$ is a monotonically increasing function of $\rho $.

In the next subsection, we exploit these results to find the current-density
relationship and explore some consequences.

\subsection{Currents and velocities}

In MTP, an occupied stack ($f>0$) contributes one (or two) hop depending on whether the next stack is empty (or occupied), the probability of which is associated with $\left(
1-f\right) $ (or $f$). Within our approximate scheme, the total contribution
is $f\left[ \left( 1-f\right) +2f\right] =f\left( 1+f\right) $. Together with the $H/L$ factor to scale from MTP to AEP, we
conclude that the mean field approximation for the particle current is 
\begin{equation}
J_\text{AC}^\text{MTP-MF}\left( \rho \right) =\left( 1-\rho \right) f\left( 1+f\right)
\label{J-ACMFA}
\end{equation}
This simple approximation provides an excellent prediction, with 
{\it no} fit parameters. The solid blues line in Figs.~\ref{fig:jac} and \ref{fig:curve} show its remarkable agreement with our simulation data.

Recall $\rho \left( 1-\rho ^{2}\right) $, the ``na\"{\i}ve
mean field approximation'' for $J\left(\rho \right) $, which is far from
simulation data (see Fig.~\ref{fig:jac} except near $\rho =0,1$). What is the difference between
the two approaches - that one performs so much better? Comparing Eq.~(\ref%
{J-ACMFA}) with $\rho \left( 1-\rho ^{2}\right) $, we see the difference to
be simply $f\left( 1+f\right) $ instead of $\rho \left( 1+\rho \right) $.
Since $f$ is an estimate of the probability of successive holes {\it %
not} being nearest neighbors, the failure of $\rho $ here implies that it
underestimates much of the {\it clustering} of the particles,. The remedy
provided by MTP can be cast as a simple and intuitive picture, for which
we coin the phrase ``abhorrence of empty
stacks.'' Since an empty stack can be filled by hops from two
(upstream) stacks, it is far less likely to remain empty, compared to the
situation in the ordinary TASEP. It is easy to check that $f-\rho $ is positive for $\rho
\in \left( 0,1\right) $ and peaks at $\thicksim 0.12$ (around $\rho
\thicksim 0.57$). In other words, $f$ is more successful at accounting for
the scarcity of empty stacks, consecutive holes in AEP language (at moderate densities).

Now let us consider ``velocity.''
There are two notions of velocity associated with $J\left( \rho \right) $. One is 
the average \textit{particle velocity}, $v=J/\rho $. From Eq.~(\ref{J-ACMFA}), we obtain a simple
expression:
\begin{equation}
v^\text{MTP-MF}=J/\rho= \left( 1-f^{2}\right)\left( 1+f\right)\label{v-ACMFA}
\end{equation}%
Since particles never hop backwards, this $v$ is necessarily positive,
approaching $1$ (or $0$) in the limit $\rho \rightarrow 0$ (or $1$). 

The other
notion of velocity\ is $\partial _{\rho }J$. Similar to the \textit{group velocity}
for waves, $\partial _{\rho }J$ is sensitive to collective behavior such as
the motion of fluctuations or disturbances. Thus, it is {\it negative} at
high densities, corresponding to holes moving ``backwards'' (the simplest case being the single-hole
system). An unusual and notable feature of AEP is ``cooperative motion'' \cite{AS00,Gabel10,GR11}. In an ordinary TASEP,
adding a particle to the system always reduces both $v$ and $\partial _{\rho }J$%
. In AEP, by contrast, there is a regime in which particles ``cooperate'' and move faster when more are present: Namely both $\partial _{\rho }v$ and $\partial _{\rho }^{2}J$ can be
positive. Both simulation data and $\partial
_{f}v^\text{MTP-MF}=\left( 1+f\right) \left( 1-3f\right) $ display such a regime.
Similarly, this behavior is also present in $\partial _{\rho }^{2}J$,
studied in  \cite{AS00,Gabel10,GR11} and easily discerned as curvature shown in Fig.~\ref{fig:curve}. Given the agreement in Fig.~\ref{fig:jac}, it
is not surprising that the MTP-MF is also successful at
predicting the phenomenon of cooperative motion.

Additionally, let us comment on two other features associated with the current $%
J_\text{AC}^\text{MTP-MF}\left( \rho \right) $. It is clear from Eq.~(\ref{f^2}) that, at $%
\rho =0.5$, we have $N=H$ and so $f\left( 1+f\right) =1$. Inserting this
result into Eq.~(\ref{J-ACMFA}), we conclude that $J_\text{AC}^\text{MTP-MF}\left( 0.5\right)
=0.5$. This is larger than simulation data by about $1\%$, a difference
which may be attributed to either statistical errors in the data or MTP-MF's
failure to capture certain correlations. Since our system does not obey
explicit particle-hole symmetry, it is unclear if $J=0.5$ is merely a
curious coincidence or an indication of some hidden symmetry. Of course,
this is the ``maximal'' value of $%
2J_\text{TASEP} $, but again, this coincidence may also be accidental. Finally,
we note that both the data and $J_\text{AC}^\text{MTP-MF}$ exceeds $2J_\text{TASEP}$ for much
of $\rho >1/2$, both peaking around $\rho \cong 0.54$. Apart from the
mathematical analysis, we have not developed a good intuitive picture for
explaining these observations.

\subsection{Cluster size distribution}
\begin{figure}[h]	
	\begin{center}
	\includegraphics[height=6cm,width=8.5cm]{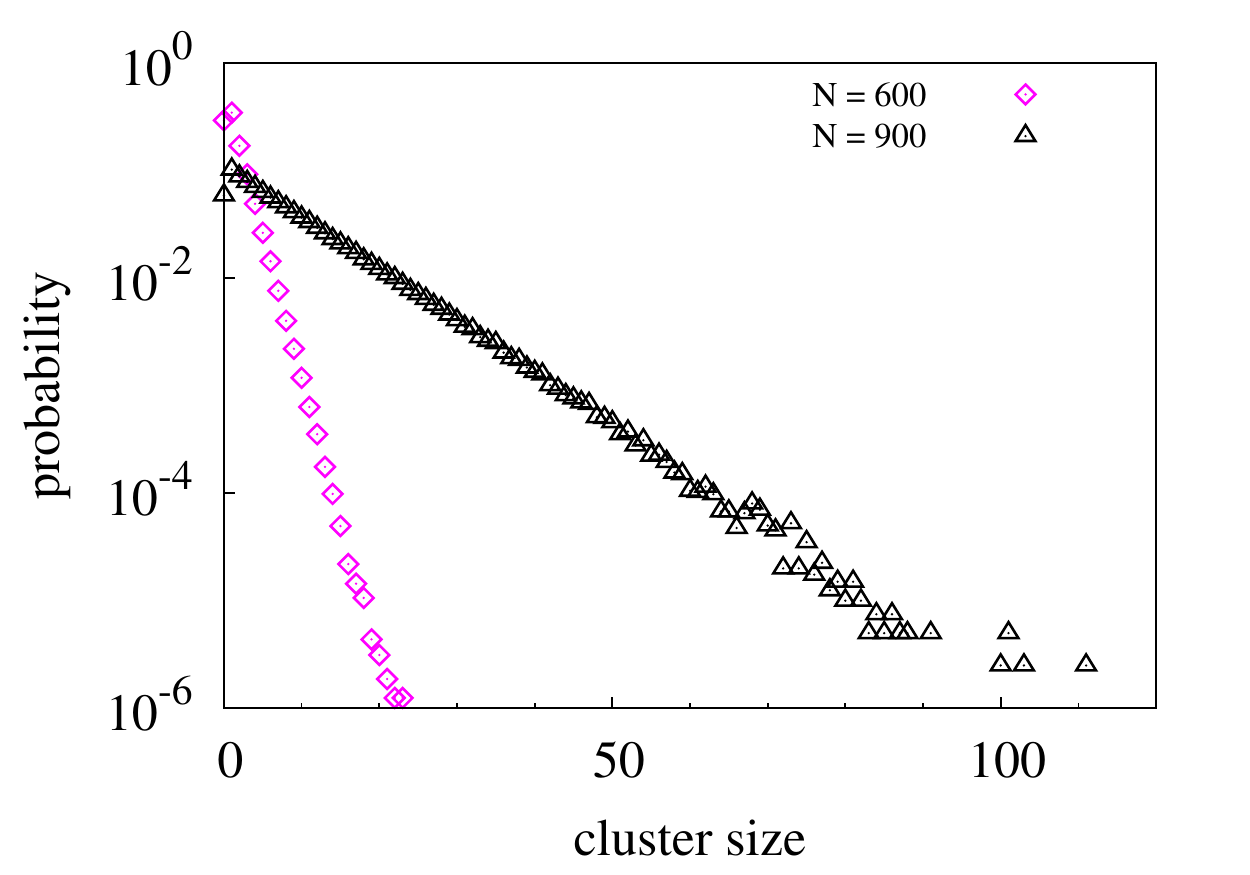}	
	\end{center} 
	\caption{(Color online) Cluster size distributions in AC phase. $\ell_\text{max} = L =1000$ with $N=$ 600 (magenta diamond) and 900 (black triangle), respectively. Theoretical predictions are not shown explicitly for clarity. The agreement between them and data are comparable to those in Figs.~\ref{fig:jac} and \ref{fig:curve}. }
	\label{fig:ac_csd}
	\end{figure}

Exploiting the MTP-MF further, we can compare the result in Eq.~(\ref{P*ell}) with
measurements of the cluster size distribution (CSD). In Fig.~\ref{fig:ac_csd}, we
show simulation data for $N=600$ and 900, both with $\ell _{\max }=1000$. It is clear that the distributions are consistent with
exponentials. Further, if we construct ratios of the successive values, we
find the average over a range (where the scatter of the data is small) to be 
$0.556$ and $0.895$, respectively. For comparison, using Eqs.~(\ref{eta}, \ref%
{f-eta}) to compute $f^{2}$, we find the values being $0.519$ and $0.895$, respectively. Another point for comparison is $P^{\ast}\left(0\right) $, for
which the simulation data provide $0.291$ and $0.0582$ versus $0.279$ and $0.0540$ from Eqs.~(\ref{P*0}, \ref{eta}, \ref{f-eta}), respectively. Such agreement leads us to conclude that the MTP-MF approach
indeed captures the essence of the AC phase.

\section{Unit Velocity Phase: Perspectives from both AEP and MTP\label{sec:UV}}

In this section, we focus on the UV branch, in which
the system is inhomogeneous and exhibits an approximately half-filled region
in coexistence with a fully occupied domain. The presence of this phase
depends crucially on having a moderate $\ell _{\max }$. The most remarkable
feature of this phase is that the average particle current is exactly $%
(1-\rho) $, shown in Fig.~\ref{fig:juv}. In other words, the average speed of the holes (or the
``disturbances'' ) is precisely $1$, {\it independent }of particles being added to, or removed from, the system
(until a phase boundary is reached). In both AEP and MTP
representations, there exist simple descriptions which provide an intuitive
and appealing picture for why such an unusual state can persist. The
following subsections are devoted to each of these perspectives. Before
presenting the details, note that such a state spontaneously breaks the
translational symmetry underpinning the dynamics. Interestingly, this broken
symmetry is manifested in slightly different forms in the two
representations. We will comment on this difference in each of the
subsections below.

\subsection{A ``hole train'' in AEP}

In the language of the original exclusion process, this phase is best
understood if we focus on how the holes move. When we choose the particle at
site $k-1$ and find that it can hop to a hole at site $k$, we can regard
this process as choosing the hole and exchanging it with the partner
particle. Next, we should ask if there is a (particle) cluster of length $%
\ell $ occupying the sites from $k+1$ on. When $\ell \in \left[ 1,\ell _{\max }%
\right] $, we also move the hole at $k+\ell +1$ to $k+\ell$. In other words, when a
hole moves, it will pull the next hole (``downstream''), provided the gap between them lies in $\left[
1,\ell _{\max }\right] $. We should remind the reader that the second hole
does not pull a third one.

Since this phase is present only in the high density regime, it is natural
to first consider systems with the lowest values of $H$. These provide
us the necessary picture to understand the existence of a UV phase. A system
with a single hole evolves trivially: The hole ``swims
upstream'' with unit average velocity, since every attempt
to move it is successful. The $H=2$ case is somewhat more interesting.
Deferring the details to Appendix \ref{app:1}, we state the principal results
here. It is straightforward to enumerate, for any $L$ and $\ell _{\max }$,
all possible stationary states, through which the important role played by $%
\ell _{\max }$ is revealed. Most significantly, for moderate $\ell _{\max
}/L $, the two holes form a {\it tightly bound} pair. To be precise, the gap
between them can be either $0$ or $1$, with equal probability. Let us
emphasize that this is an absorbing state. Starting with any initial
separation, the two holes will eventually drift together, form the bound
state, and never become unbound thereafter. Since the ``leading'' hole always moves when it is chosen, the pair
moves with unit average velocity.

The first non-trivial case is $H=3$. For a finite system with arbitrary $%
\ell _{\max }$, an exact solution is not yet available. Nevertheless, we can
gain some insight into the UV behavior through an exact result in the limit 
$L\rightarrow \infty $ followed by $\ell _{\max }\rightarrow \infty $ (details in Appendix \ref{app:2}). Here,
the leading pair remains tightly bound, as the third hole cannot affect
the leading hole. Denoting the gap between the second and the third hole by $m$, we know that the third hole can trail behind the second by $%
m\geq 0$, so that the complete description of the system lies in the
following probabilities. Let $p_{0, 1}\left( m\right) $ be the probability that the gap between the first hole pair is $0$
or $1$, with the third hole trailing by $m$ sites.
We illustrate examples of $p_{0}\left( 5\right) $ and $p_{1}\left( 4\right) 
$ in Fig.~\ref{fig:3h}. As shown in Appendix \ref{app:2}, we find (apart from $m=0,1$)%
\begin{equation}
p_{0,1}\left( m\right) \propto \zeta ^{m}
\end{equation}
where $\zeta =2-\sqrt{2}\cong 0.586$. In other words, the third hole is
``loosely'' bound, with an exponential tail
of a characteristic length $\mu \equiv -1/\left( \ln \zeta \right) \cong
1.87$. The intuitive picture is clear: The first pair advances with UV,
with the third hole being ``pulled along,''
at a typical distance $\mu $ behind the second. It is natural to label such
a triplet a ``hole train'' with the first
pair being the ``engine''\cite{Dong12}.

	\begin{figure}[h]	
	\begin{center}
	\includegraphics[height=2.8cm,width=8.2cm]{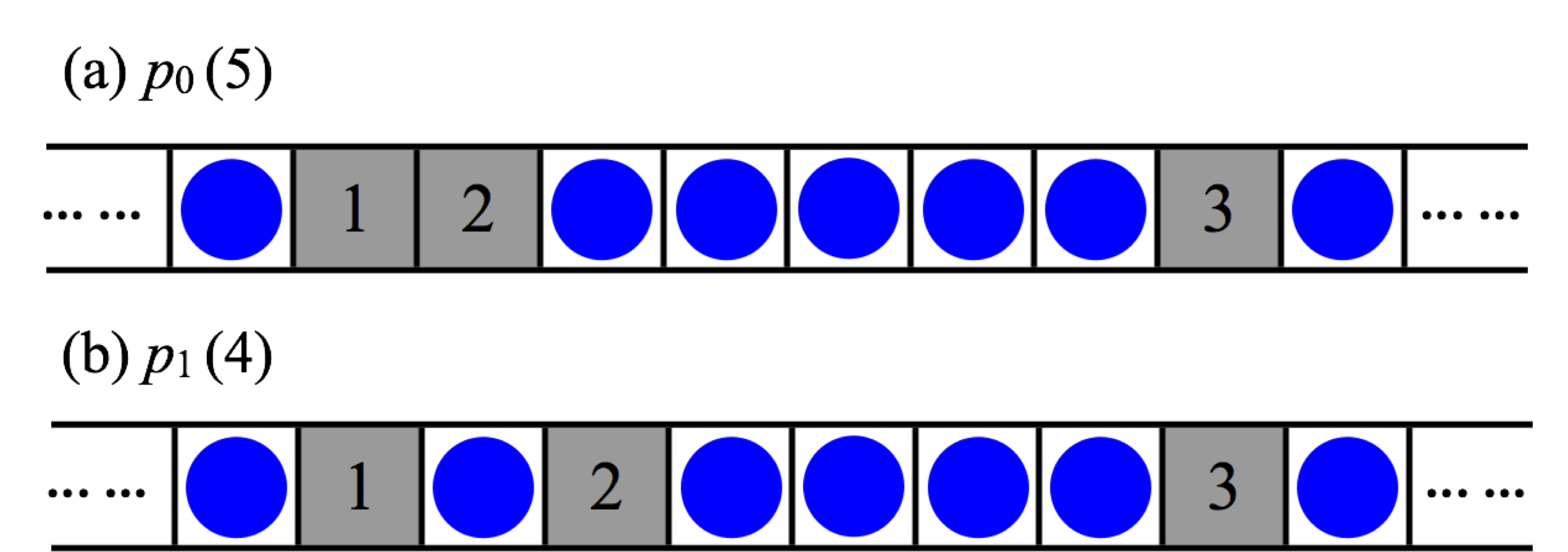}	
	\end{center} 
	\caption{Sketch of a ``hole-train'' with $H=3$: (a) Two leading holes are consecutive with the third trailing by $m=5$. The probability of this scenario is denoted as $p_0(5)$. (b) Two leading holes have one particle in between, with the third trailing by $m=4$, denoted as $p_1(5)$.}
	\label{fig:3h}
	\end{figure}

Continuing this line of thought, we see that, in the limit posed above, the
first pair will always form an ``engine''
which advances with UV, unaffected by how many holes trail behind them.
While we do not have exact solutions for the general case, we can argue why
the holes should be bound and the image of a train is quite reasonable. In
particular, consider the last hole -- the ``caboose'' in the parlance of freight trains -- and the penultimate
hole, which we will refer to as ``X.'' If X
is chosen and moves, it will pull the caboose along (provided the gap
in-between is non-zero). The gap remains the same. On the other
hand, if the caboose moves, the gap decreases. Only when X is
pulled along by the hole upstream from it does the gap increase. Although
these considerations appear to imply that the gap between penultimate and final holes performs a typical random
walk, we point out that the entire train length does not increase even when
this gap increases. Thus, we argue that this mechanism can ``hold the train together.'' Accepting this scenario, we see
that adding or removing holes to the system merely changes the total length
of the train. Meanwhile, since the engine advances with UV, the whole train
also moves as such, leading to an $H$-independent velocity. In the following
section, we present a more tenable and quantitative argument for the existence of this hole train, as
well as an estimate of its average density ($\thicksim 0.5$) in the language of MTP.

Before proceeding, let us comment on two other aspects of the AEP
perspective. First, although a typical snapshot of our system clearly
violates translational invariance, this symmetry is restored quite quickly ($%
O\left( L\right) $ MCS) since the hole-train moves at UV. To be precise, if
we measure the occupation at a specific site, it will settle at $\rho $
within such times. If, on the other hand, we tag a particular hole and
measure the occupation in one of its nearest neighbor sites, then the
results will expose the inhomogeneity inherent in the system. Restoration of
the symmetry (in finite systems) would take much longer than $O\left(
L\right) $ MCS, a subject well beyond the scope of this work. Second, in a
finite {\it periodic }lattice, the average distance between the caboose and
the engine, $\Delta $, is finite. For systems with $\Delta \gg \ell _{\max }$
, the caboose does not affect the (lead hole in the) engine. Thus, the
integrity of the engine remains intact and moves the entire train with UV.
As holes are added or removed, $\Delta $ becomes smaller or larger,
respectively. When enough holes are added, or if $\ell _{\max }$ is raised,
then $\Delta $ can approach $\ell _{\max }$, the caboose can destroy the
engine, and the train can become unbound. In a nutshell, this is the
mechanism for the transition from UV to AC, as the overall density is lowered (e.g.,
Fig.~\ref{fig:juv}). We will return to this picture in Section \ref{sec:trans}.

\subsection{ Condensation in MTP}
As for the AC phase, the MTP representation provides us with a more
quantitative picture. The solid particle cluster plays the role of the
condensate in non-trivial ZRP's. With a finite (and small enough) $\ell
_{\max }$, it is possible for one stack to contain more balls than $\ell
_{\max }$. Such a stack can gain balls in two possible ways, much like how an
empty stack can be filled above. Meanwhile all stacks can lose a ball in just
one way. Thus, the number in this stack will grow, until a steady state is
reached. It is natural to refer to this behavior as ``condensation'' and this stack (or the balls in this stack) as
the ``condensate.'' To make contact with
the previous subsection, note that there are $\Delta $ balls in the condensate stack.
Needless to say, each of the remaining $H-1$ stacks is likely to hold very few
balls. In the language of ZRP, the state of the other stacks is referred to as
``a fluid.'' Here, we recognize them as the
hole train. Further, this picture allows us to appreciate better why the
hole train remains bound. First, the train length is monotonically
related to the fluid density, becoming longer/shorter when the condensate
loses/gains balls. Second, the only way to redress the imbalance (two
gains {\it vs}. one loss) for the condensate is when the fluid density
remains relatively low, i.e., a set of stacks with few balls in each. This
scenario corresponds to a bound train.

Turning to a more quantitative description of the steady state, we denote the occupation probability within the train by $f_\text{train}$. A good estimate for it comes from the balance of the gain/loss contributions
from the condensate, namely%
\begin{equation}
f_\text{train}+f_\text{train}^{2}=1
\end{equation}
The predicted value, $f_\text{train}=\left( \sqrt{5}-1\right) /2$, is not very
illuminating. Instead, by comparing with Eq.~(\ref{f^2}), we find a more
insightful relation:
\begin{equation}
H_\text{train}=N_\text{train}
\end{equation}%
namely, $\rho _\text{train}=0.5$, implying a train length is $2H_\text{train}$. This
result also indicates that the typical distance from one hole to the next in
the train is around $2$, a picture entirely consistent with the result $\mu
\thicksim 2$ in the $H=3$ case.

Meanwhile, since the train consists of $H-1$ stacks, we arrive at $%
H_\text{train}\cong H$. Further, we have $N_\text{train}+\Delta =N=L-H$ so that the
size of the condensate is given by:
\begin{equation}
\Delta \cong L-2H=L\left( 2\rho -1\right)   \label{Delta}
\end{equation}

\begin{figure}[h]	
	\begin{center}
	\includegraphics[height=4.5cm,width=9cm]{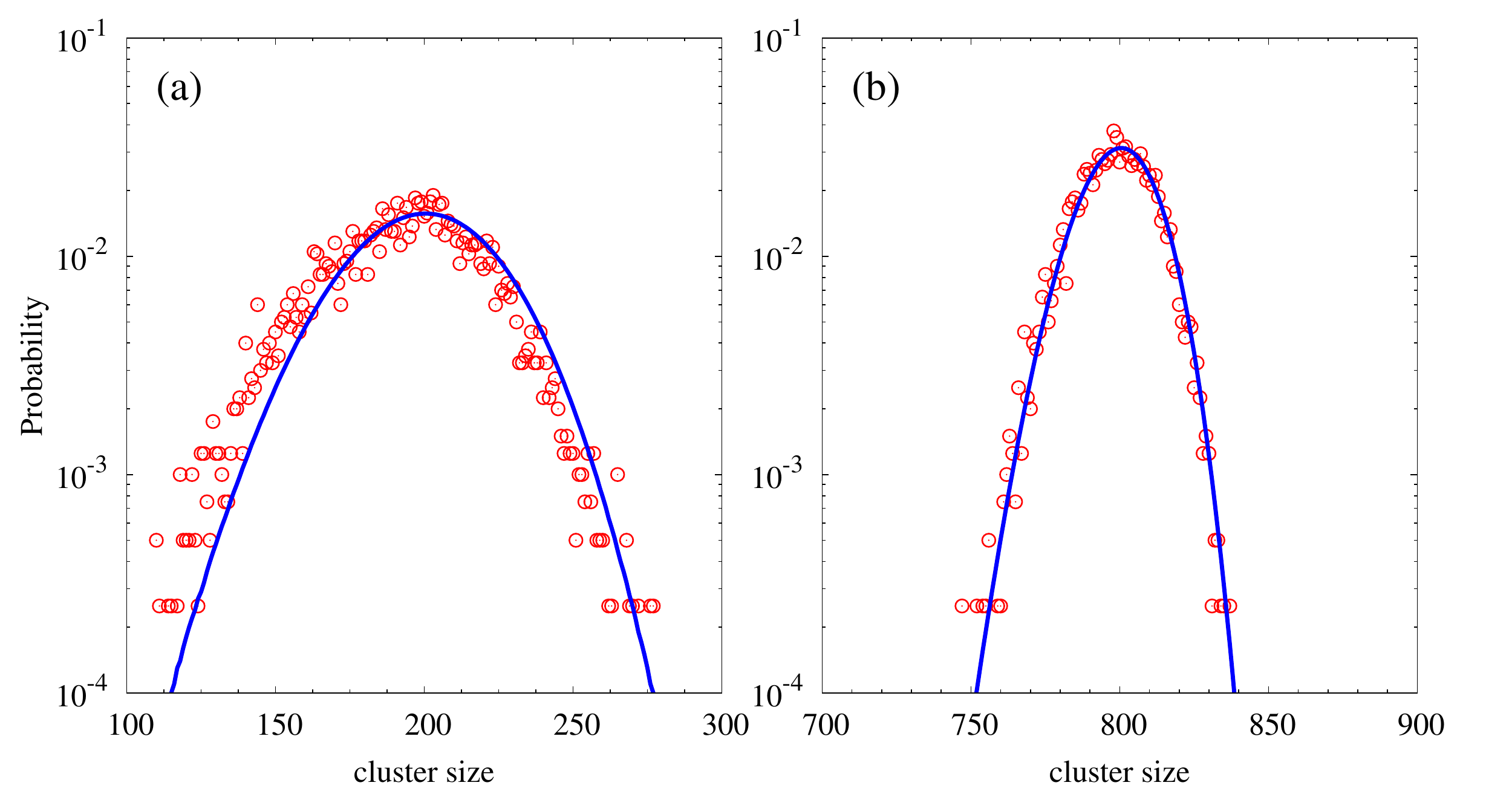}	
	\end{center} 
	\caption{(Color online) Cluster size distributions in UV phase. $\ell_\text{max}=100$, $L=1000$. Circles are simulation data and the solid line is the result from MTP-MF. (a) $N=600$. (b) $N=900$.}
	\label{fig:uv_csd}
	\end{figure}

All these predictions are borne out relatively well in simulations. For
example, in Fig.~\ref{fig:uv_csd}, we show the CSD's for $N=600$ and $900$ with $\ell_\text{max}=100$.
Clearly, the condensate sizes are seen to fluctuate around $200$ and $800$
respectively, as predicted by Eq.~(\ref{Delta}). 
The properties of the fluid/hole-train, as revealed by the small clusters not shown in Fig.~\ref{fig:uv_csd}, are essentially the same in both cases. For small clusters the distribution indeed decays exponentially,  with ratios of the successive values being $0.406$ and $0.412$, respectively. These are
somewhat higher than $f_\text{train}^{2}\cong 0.382.$
Given that our theory is based on a mean field approximation, we speculate that
the difference are due to non-trivial correlations, the study of which is
beyond the scope of this work.

So far, the analysis is focused on the average behavior of the fluid and the
condensate. Since our approach considers the single-stack occupation, we can
apply it to the condensate and exploit a self-consistent way to predict $%
P^{\ast}_\text{con}\left(\Lambda \right) $, the probability for the condensate to
have $\Lambda $ balls in the steady state. Note that, unlike $\Delta $, $%
\Lambda $ is a variable here. In such a configuration, the fluid has only $%
N-\Lambda $ balls, which allows us to estimate $\tilde{f}$ (the occupation
probability of a stack in the fluid) as a function of $\Lambda $. Using Eqs.~
(\ref{eta},\ref{f-eta}), we find $\tilde{f}\left( \Lambda \right) =\exp
\left\{ -\sinh ^{-1}\frac{H-1}{2\left( N-\Lambda \right) }\right\} $ while $%
\tilde{f}\left( \Delta \right) $ is just $f_\text{train}$. We can now use $%
\tilde{f}$ to estimate the rate at which the fluid supplies a ball to the
condensate, namely, $\tilde{f}^{~2}+\tilde{f}$. Denoting this rate by%
\begin{equation}
g\left( \Lambda \right) \equiv \tilde{f}^{~2}+\tilde{f}
\end{equation}%
we find an expression similar to Eq.~(\ref{P01})%
\begin{equation}
g\left( \Lambda \right) P^{\ast}_\text{con}\left( \Lambda \right) =P^{\ast}_\text{con}\left(
\Lambda +1\right) ,
\end{equation}%
since the condensate loses at unit rate. It is straightforward to check that 
$g\left( \Lambda \right) $ is a monotonically decreasing function and is
unity at $\Lambda =\Delta $. Thus, $P^{\ast}_\text{con}\left( \Delta +1\right)
=P^{\ast}_\text{con}\left( \Delta \right) \equiv \hat{P}$ are the peak values of the
distribution and can be conveniently used to start the recursive evaluation
of two sequences: $P^{\ast}_\text{con}\left( \Delta +1+k\right) $ and $P^{\ast}_\text{con}\left(
\Delta -k\right) $ with $k>0$. Furthermore, even though $g$ appears to depend on
both control parameters $\left( H,N\right) $, it actually is a function of a
single (shifted and scaled) variable%
\begin{equation}
\xi \equiv \frac{\Lambda -\Delta }{H-1}=\frac{\Lambda -N+H}{H-1}.
\end{equation}%
For completeness, we provide the explicit expression: 
\begin{equation}
g\left( \Lambda ;H,N\right) =1-\frac{\xi }{1-\xi }\exp \left\{ -\sinh ^{-1}%
\frac{1}{2\left( 1-\xi \right) }\right\} 
\end{equation}%
Now, we can express $\ln P^{\ast }_\text{con}$ as a sum over $\ln g:$%
\begin{eqnarray}
\ln P^{\ast}_\text{con}\left( \Delta +1+k\right)  &=&\ln \hat{P}-\sum_{{}}\ln g\left(
\xi \right)   \label{lnP*+} \\
\ln P^{\ast}_\text{con}\left( \Delta -k\right)  &=&\ln \hat{P}+\sum_{{}}\ln g\left(
\xi \right)   \label{lnP*-}
\end{eqnarray}%
where the sums run over $\xi $ being integer multiples of $1/\left(
H-1\right) $, up to $\pm k$. Although we cannot evaluate this sum, we can
extract its properties for large $H$ and $k$ of $O\left( 1\right) $ (i.e.,
small $\xi $'s). To leading order, the resultant $\ln P^{\ast}_\text{con}$ is a
function of $k^{2}/H$. In other words, the condensate size distribution (at
this level of approximation) is universal in the following sense: Although its
explicit dependence is $P^{\ast}_\text{con}\left( \Lambda ;N,H\right) $, it can be
cast in scaling form, $P^{\ast}_\text{con}\propto \Phi \left( x\right) $, where $\Phi $
is a universal function of the scaled variable
\begin{equation}
x\equiv \frac{\Lambda -N+H}{\sqrt{H}}.
\end{equation}%
Such behavior is similar to Gaussian distributions, which are universal
apart from a displacement and a rescaling. A detailed study of $\Phi $ is in
progress and will be reported elsewhere. Here, let us present the
numerical results from Eqs.~(\ref{lnP*+},\ref{lnP*-}) for the cases above.
The agreement with the two data sets are again remarkably good (Fig.\ref{fig:uv_csd}).
We note the slight discrepancies in the $\Delta =200$ case, and believe that
they are the consequences of the fluid section being longer ($H$ being $400
$ instead of $100$). Surely, for systems with larger fluid components, the
fluctuations therein will be more serious. Obviously a careful study and
analysis of such fluctuations and correlations will be necessary if the goal
is to go beyond mean field theory.

To summarize, we see that a quantitatively coherent picture of the hole train
emerges when viewed in the MTP representation. Here, the hole-train
corresponds to the fluid, while the solid cluster in the rest of the lattice
corresponds to the condensate. Translational symmetry in the MTP is
spontaneously broken, as the condensate resides in one of the $H$ stacks.
However, unlike in the AEP picture, this condensate does not move with UV
from stack to stack. Symmetry restoration must proceed by evaporation and
re-condensation. Such a process is expected to take considerably longer than
the $O\left( L\right) $ MCS in the AEP representation, while its detailed
nature is being investigated \cite{DMprivate}. The results here allow us to take the
thermodynamics limit: $L,N,H,\ell _{\max }\rightarrow \infty $ with finite

\begin{equation}
\rho =N/L;~~\lambda =\ell _{\max }/L.
\end{equation}%
Provided $\lambda <2\rho -1$, it is possible for a hole train to form,
occupying a finite fraction ($2-2\rho $) of the ring. The gap between the
engine and the caboose, corresponding to the size of the condensate, fills
the remaining fraction: $2\rho -1$. This result clearly implies that the UV
phase cannot exist for $\rho <1/2$.

\section{Transitions between AC and UV\label{sec:trans}}

In this section, let us consider the transition between the two phases and
map out a phase diagram in the $\rho-\ell _{\max }$ plane. First, note
that the thermodynamic limit cannot be studied rigorously, especially since
the exact steady state distribution, $P^{\ast }\left( {\cal C}\right) $, is
not known. If this limit does not exist, then the standard term ``phase'' should be used with some caution.
Second, the standard approach to phase transitions involves taking this
limit with the {\it stationary} state, i.e., the limit $t\rightarrow \infty $
is taken first, while simulations are based on running finite systems ($%
L<\infty $) for finite times ($\tau <\infty $). Therefore, we can only make
some estimates and offer some rough arguments here. Obviously, more
convincing conclusions can be drawn from a thorough finite-size scaling
analysis (See e.g. \cite{Cardy88}), a task beyond the scope of this paper. Finally, we should
comment on the order parameter, i.e., how we characterize the phases. By
studying mainly the current $J$, we have implicitly chosen (the operators
in) Eq.~(\ref{AEPJ}) here. Yet, as discussed in the
previous two sections, the phases may be better characterized by the
presence/absence of a macroscopic cluster. Thus, another possibility is to
study the distribution of the size of the largest cluster, which we denote
by $Q\left( s\right) $. Deep in the AC phase, $Q$ should be similar to $%
P^{\ast }\left( \ell \right) $ for large $\ell $. From Eqs.~(\ref{P*ell},\ref%
{f-eta}), we therefore expect $Q\left( s\right) \rightarrow e^{-2\eta s}$. On
the other hand, deep in UV, this cluster is the caboose-engine gap or the
condensate, so that $Q\left( s\right) $ is just $P^{\ast }_\text{con}\left(
\Lambda \right) $. Indeed, most of our theoretical arguments for the phase
transition presented below will be based on the properties of $Q$.

Since we are dealing with non-equilibrium steady state, there is no widely
accepted notion of a ``free energy'' even
if we managed to find an explicit $P^{\ast}\left(\mathcal{C}\right) $. Thus,
we cannot follow the standard route, defining a first order transition
through a jump in its derivative. Alternatively, we can define such a point
dynamically, given that our system is formulated as a stochastic process. A
reasonable choice is, for example, that set of control parameters with which
the system settles for equally long periods in each of the phases while
switching between them occasionally (``tunneling'' ). If computer power/time is unlimited, we can
measure $s\left( t\right) $ for arbitrarily long periods and compile a
histogram for $Q\left( s\right) $. If our analytic power is strong enough,
we can access the exact steady state $Q$. For a range of parameters, $Q$
should be sharply bimodal, allowing us to locate special points where the
modes are equally probable. However, as both approaches are quite limited at
present, we can offer only rough estimates and reasonable arguments for a
``phase diagram'' here. Three regimes are
expected to be present: pure AC, pure UV, and ``mixed'' (AC+UV). The best way to characterize these regimes
is through lifetimes. Specifically, deep in the pure regimes, the system settles relatively quickly into
one phase, regardless of initial conditions. By contrast, in the mixed
regime, once it settles into AC or UV (typically through judicious choice of
initial conditions), that state can persist for extraordinarily long times.
In particular, it is possible for these lifetimes to scale exponentially
with the system size, $L$. In that case, such a regime rightly deserves the
label ``bistable'' 
\footnote{In the $H$-$T$ plane of the standard Ising model (with say, periodic BC),
such a regime is just a line: $H=0$ and $T$ below criticality. By contrast,
Toom has shown that \cite{Toom80}, if driven out of equilibrium in a certain way, this
line will expand into a finite-area region symmetric about $H=0$. 
}.

We emphasize that there are three independent control parameters in
the simple AEP. They can be $L$, $\rho $, and $\lambda $, or $N$, $H$,
and $\ell _{\max }$, for example. Also, a variety of ``thermodynamic limits'' can be taken, depending on the order
that different quantities are sent to infinity. In addition, for
simulations, the initial condition and length of runs will be important to
consider. Enumerating all possibilities is exceedingly difficult, if not
impossible. Here, we will focus mainly on the parameters used in our
simulations ($L=1000$ and a wide range of $\rho $ and $\ell _{\max }$) and
provide some arguments from our analysis for other situations.

\begin{figure}[h]	
	\begin{center}
	\includegraphics[height=6cm,width=8.5cm]{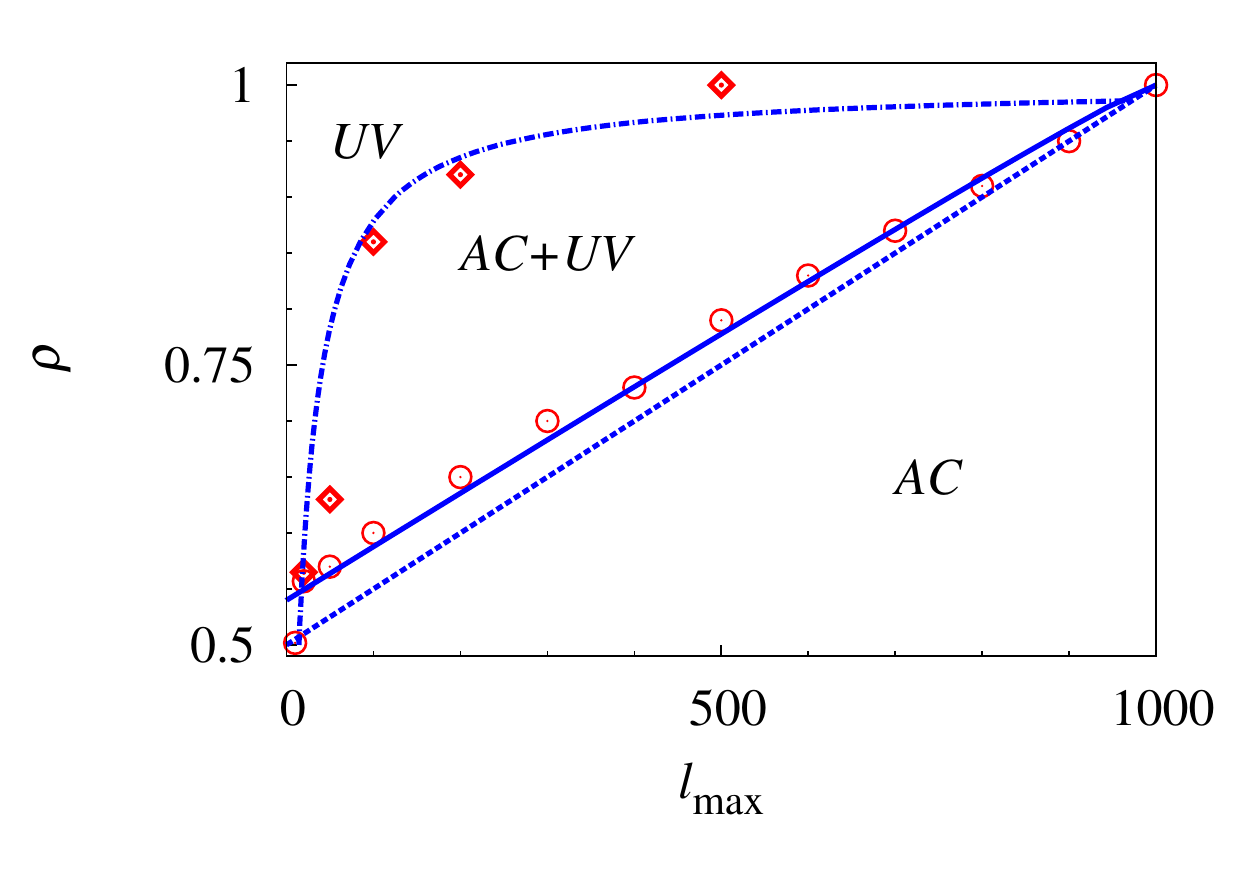}	
	\end{center} 
	\vspace{-0.7cm}
	\caption{(Color online) Phase diagram in $\rho$-$\ell_\text{max}$ plane. Symbols are from simulations. Phase boundaries are results from Eqs.~(\ref{rhoc1},\ref{rhocA},\ref{eta_cU}) (dash, solid and dot-dash), respectively.}
	\label{fig:phase}
	\end{figure}

It is clear that if we use a totally inhomogeneous system (all particles
clustered together) as an initial condition, then we are likely to find the
pure UV regime, as well as to explore the boundary between the mixed and
pure AC regimes. Indeed, all the simulation data presented above are
collected under these conditions. As indicated in the previous section, in AEP, UV is associated with a 
``hole-train'' of typical length $2H$, which will be
destroyed if the caboose wanders within $\ell _{\max }$ of the engine. Thus,
UV is unstable if $\ell _{\max }\geq L-2H$. This provides an estimate for
the critical density associated with the boundary with the pure AC phase \cite{Dong12}:
\begin{equation}
\ell _{\max }=L\left( 2\rho _{cA}-1\right)
 \label{rhoc1}
\end{equation}%
To be explicit, a system with $\rho <\rho _{cA}\cong \frac{1+\lambda }{2} $ will settle into AC only.
This boundary is shown as the dashed (blue) line in Fig.~\ref{fig:phase} (with $
L=1000$). Although the agreement with data (red circles) is reasonably
acceptable, this estimate can be improved by incorporating some
fluctuations. In MTP, the caboose-engine gap appears as the condensate
size, $\Lambda $, which fluctuates around $\Delta $. Thus, $\Lambda $ can
reach $\ell _{\max }$ with a small probability, $P_\text{con}^{\ast }\left(
\ell _{\max }\right) $, even if $\Delta $ may not be near $\ell _{\max }$.
Now, we may argue that, in a run of $\tau $ MCS, rare events with
probability $1/\tau $ can occur. Exploiting this connection and
approximating the scaling form for $P_\text{con}^{\ast }\left( \Lambda \right) $
by a Gaussian, we see that the UV can become unstable if $\Delta -\ell
_{\max }\thicksim O\left( \sqrt{H\ln \tau }\right) $. Inserting $\tau
\sim 10^{6} $ used in our simulations, we find that Eq.~(\ref{rhoc1}) is slightly
modified. As an illustration, we plot
\begin{equation}
\ell _{\max }=L\left(2\rho _{cA}-1\right) -\sqrt{L(1-\rho _{cA})\ln \tau }.
\label{rhocA}
\end{equation}%
Shown as the solid (blue) line in Fig.~\ref{fig:phase}, it is arguably an improvement.
If this result is upheld in a more rigorous analysis, we may conclude
that if the thermodynamic limit is taken first, there is a non-trivial
region in the $\rho $-$\lambda $ plane associated with AC+UV bistability,
while Eq.~(\ref{rhoc1}) marks its border with the pure AC regime.

Next, we explore the stability of the AC phase. In simulations, this phase
will be more favored by distributing particles uniformly on the lattice
initially. In this manner, we expect to find another boundary, beyond
which the system never settles in AC. Deferring a systematic investigation,
we simply provide a few examples in Fig.~\ref{fig:phase} (red diamonds).
Theoretically, our approach is similar to the one above: What is the
probability for the largest cluster to reach $\ell _{\max }$ particles?
Within our approximate scheme, this is given by $P^{\ast }\left( \ell _{\max
}\right) $, and using Eqs.~(\ref{P*ell},\ref{eta},\ref{f-eta}), we arrive at $\exp \left(
-2\eta \ell _{\max }\right) $. Applying the connection to runs of length $%
\tau $, we obtain%
\begin{equation}
2\eta _{cU}\ell _{\max }\thicksim \ln \tau  \label{etacU}
\end{equation}%
where $\eta _{cU}$ is related to the critical density associated with the
boundary UV regime in Eq.~(\ref{eta}): 
\begin{equation}
\sinh \eta _{cU}=\frac{1-\rho _{cU}}{2\rho _{cU}}
\label{eta_cU}
\end{equation}%
The resultant is also plotted in Fig.~\ref{fig:phase} (dot-dash blue line).
While the discrepancies between this estimate and data is larger than those
above, we may conclude that this approach is a viable first step. In
particular, we believe that a major difference between these two cases lies
in the following. For a system in UV to tunnel to AC, the condensate stack
needs to wander all the way down to $\ell _{\max }$. We can formulate this problem of
finding the lifetime as a first passage time of a single walker arriving at
a particular destination. By contrast, in the reverse process, tunneling
from AC to UV requires only one of the stacks to wander up to $\ell _{\max }$%
, corresponding to finding the first time that any one of the $H$ walkers
arrives at the destination. Clearly, the latter problem is more complex,
especially since the walkers are not entirely independent. To improve on Eq.~(%
\ref{etacU}) will be a worthy next step.

To summarize, we presented a plausible phase diagram associated with the
discontinuous transitions observed in simulations, consisting of three
regimes. In two of these regimes, the system appears to evolve to a unique
steady state: AC or UV. In between, our simulations show that the system can
settle into either state, depending on, e.g., initial conditions. We
conjecture that our system supports the phenomenon of bistability, namely,
the time scales for switching between these states grow exponentially with
they system size. In other words, we expect the behavior here to resemble
that in equilibrium systems with long range interactions (See., e.g., \cite{Kac59} and more recently \cite{Grewe77, Lee01}). To prove or disprove this
conjecture will likely be accomplished through careful observations of
hysteresis along with a finite size scaling analysis.

\section{Summary and outlook\label{sec:sum}}

In this article, we investigated an accelerated exclusion process (AEP) on a ring where particles hop when the neighboring site is empty, as well as  kick another one forward when joining a cluster of particles of size $\ell \leq \ell_\text{max}$. 
Through Monte Carlo simulations, we discovered that, with various choices of density $\rho$ and interaction range $\ell_\text{max}$, the system may be found in an augmented current phase or a unit-velocity phase. 
The behavior this AEP exhihibits, both dynamic and static, are much richer than the standard TASEP. 
Focusing on the steady state, we expand the findings reported in Ref.~\cite{Dong12} and seek a comprehensive theoretical framework for understanding these novel features.
The apparent inadequacy of a na\"{\i}ve mean field approach prompted us to seek alternative routes. Treating AEP as a mass transport process (MTP) of balls contained in stacks and may jump either one stack ( ``hop'' only ) or two ( ``hop and kick'' ), we provide a more intuitive picture of both phases and the transition between them. In this representation, a mean field approximation scheme is formulated to compute several key quantities. With no fit parameters, the predictions agree remarkably well with results from simulations. 

For the AC phase, we found an expression for the particle current, $J_\text{AC}^\text{MTP-MF}=(1-\rho)f(1+f)$, with $f$ being the probability of an occupied stack (equivalently, the frequency of isolated holes in AEP), given explicitly by Eqs.~(\ref{eta},\ref{f-eta}). This result enabled us to quantitatively estimate both $J$ and the ``acceleration'' in the facilitated region. Additionally, it provided an intuitive picture for the scarcity of hole pairs, which leads to more ``kicks'' and augmented currents.

Once the system is in the UV phase, $J$ is simply $(1-\rho)$ regardless of $\ell_\text{max}$, indicating the holes in the system are traveling at unit velocity. 
This intriguing result can be appreciated from the AEP and MTP representations with different insights. In the language of AEP, the system in UV is composed of a ``hole-train'' of length $2H$, led by an ``engine'' composed of a tightly bound hole-pair. 
The complement of the hole train is a cluster of particles of size $\Delta=L-2H$. 
In the language of the MTP, the hole train and cluster is, respectively, the fluid and the condensate. 
We also computed the leading term in $P_\text{con}^{\ast}$, which enabled us to understand the average sizes of the condensate as well as its fluctuations. Preliminary studies of scaling behavior and a universal distribution are encouraging and further investigations are in progress.  

Deferring charting a precise phase diagram the AC and UV phases to our further quests, we reported the essentials in the formation of condensates, thus infer the phase boundary using $\rho$ and $\ell_\text{max}$ as order parameters.  Starting as a ``solid'' (all balls in one stack) or a ``liquid'' (balls distributed through all stacks) leads the system to favor UV or AC. Various factors, including the initial conditions, affect where the system eventually settles and how long it remains, hence we conjecture the two different phase boundaries presented in Fig.~\ref{fig:phase} with support from our simulations. 

Although this study provided valuable insights into the AEP, there are many
avenues to improve on both simulations and theory, in order to advance a
better understanding of its behavior. Examples mentioned above include
explorations of the dependence on  $L$ and $\tau $, finite size scaling analysis, and
a careful study of clusters' evolution and the size distributions. On the
theoretical front, we should account for some correlations in the system,
for example, by considering the joint distribution $P\left( \ell _{1},\ell
_{2}\right) $ in the MTP. A more refined phase diagram than our Fig.~\ref{fig:phase}
would be most desirable. In particular, we may expect that the
discontinuous jumps give way to a continuous, second-order like, phase
transition. 
Subsequently, all the standard issues associated with such a transition
can be explored, from critical exponents and universality classes to scaling
and renormalization group analyses. Beyond static properties, we envisage
many interesting dynamic questions. In addition to investigations already in
progress \cite{DMprivate}, it would be instructive to study time series and power
spectra of various quantities, since they can expose the details of
correlations in time. For example, we may study microscopic currents
associated with entry and exit times from the condensate. While the latter
is expected to be simply Poisson distributed, the former may be more complex,
as it is connected to the fluctuations of the entire fluid. In particular,
the correlations of entry/exit times should provide information on
propagation of fluctuations (through the fluid).

Beyond the system studied here, there are natural generalizations, such as
having two or more particles being activated and $\ell $-dependent kicking
probabilities. Another natural generalization is the AEP with open boundary
conditions, with a variety of injection/extraction possibilities. Mapping
out the equivalent of the open TASEP phase diagram fully will be an arduous,
but rewarding task. We may introduce inhomogeneous hopping rates modeling
blockages or adsorption/desorption along the entire chain. Further afield,
we may wish to consider systems with many species, or many lanes
(``quasi-1D'' ), as well as in higher
dimensions. Yet another important task  for the future is to see to what extent the features of AEP are present in more realistic models of systems in nature that display assisted hopping. 
Finally, we hope
that AEP will be a new window for understanding not only exclusion
processes, but also non-equilibrium statistical mechanics in general.

\section{Acknowledgements}

We acknowledge insightful discussions with H. Hilhorst, K. Mallick, S.
Redner, B. Schmittmann, and J.M.J. van Leeuwen. We are
especially grateful to D. Mukamel for communicating privately their findings \cite{DMprivate}, many of which are similar to ours. This research is supported in part by the US National Science
Foundation through grants DMR-1244666 and DMR-1248387.

 \appendix
 \section{Exact solution for $H=2$\label{app:1}}
For $H=2$, the system is sufficiently trivial that we can simply enumerate
all possibilities, denoted as a pair of integers in parenthesis. In the MTP representation, we need to consider only the
number of balls in one stack, $\ell \in \left[ 0,N\right] $. The other stack
contains $N-\ell $. Furthermore, there are at most three intervals in $\left[
0,N\right] $, in which the rules are different.

If $\ell _{\max }\geq N$, then every ball moves two steps (returning to the
original stack), so that every ``interior'' (i.e., $\ell \in \left[ 1,N-1\right] $) configuration stays the same.
Meanwhile, each of the two ``boundary'' configurations decays as $2^{-t}$, since choosing the filled stack will lead to
a stationary one. Though it appears to be stationary in the MTP
representation, the AEP current is always $2$ no matter which particle-hole
pair is exchanged. It is natural, therefore, for us to give such a state the
label AC. Since every initial condition corresponds to such a state, the
system is always in AC.

If $N>\ell _{\max }\geq N/2$, then we can have a non-maximal ``interior region'' ($N-\ell _{\max }<\ell
<\ell _{\max }$) of stationary configurations, as in the previous paragraph.
If the initial condition is in this region, the system is again AC. However,
if the system starts in one of the two ``boundary
regions,'' then it will evolve as follows. Such a
configuration consists of one stack with $\ell >\ell _{\max }$ balls and the
other with $N-\ell \leq \ell _{\max }$. If the latter is occupied, then
a ball leaving the first stack will make two hops and return to the original stack.
Thus, this stack either gains a ball or remains the same, so that $\ell $
tends to drift upwards. In other words, the system performs a {\it biased}
random walk towards the boundary until it reaches the configuration: $%
\left( N,0\right) $. From there, it can only reach $\left( N-1,1\right) $.
From this point, the system jumps between these two configurations, i.e., a
stationary state we recognize as the tightly bound pair (the ``engine'') discussed in Section \ref{sec:UV} A. It
is clear (and straightforward to prove) that such a system should be labeled
by UV. Thus, a system with initial conditions in these ``boundary regions'' simply evolves towards a UV stationary
state.

Finally, if $N/2>\ell _{\max }$, the configurations in the
``interior region'' ($\ell _{\max }<\ell
<N-\ell _{\max }$) consist of both stacks having more than $\ell _{\max }$
balls. Now, balls just move from one stack to the other, so that the system
performs an unbiased random walk in $\ell $. When it reaches one of the
 ``boundary regions,'' it converts to
performing a biased random walk as above. Thus, such system will always end
in a UV stationary state.

To summarize, the $H=2$ case, though seemingly trivial, provides the
essentials of the AC and UV ``phases.'' The
simplest ``phase diagram'' emerges: A
domain in $N$-$\ell _{\max }$ plane with pure AC, one with pure UV, as well
as a third where both AC and UV can be the end state (depending on initial
conditions).

\section{Solution for $H=3$ in a special limit\label{app:2}}
Clearly there are  more possibilities for the $H=3$ case and they would be more complicated. Enumerating them and providing exact solutions in each scenario remain to be completed. Here, let us consider a special
limit, $L\rightarrow \infty $ followed by $\ell _{\max }\rightarrow \infty $. The former limit implies that the caboose cannot affect the engine, which
remains a tightly bound pair. The latter ensures that the engine can affect
the caboose, which can lag behind by an arbitrary number ($m\geq 0$) of
sites. In this scenario, we only need to consider $p_{0}\left( m\right) $
and $p_{1}\left( m\right) $, the probability that the gap between the first
pair is $0$ and $1$, respectively, with the caboose trailing by $m$ (see Fig.\ref{fig:3h}). In the
stationary state, these satisfy:
\begin{eqnarray}
\begin{split}
2p_{0}\left( m\right)  &=p_{0}\left( m+1\right) +p_{1}\left( m\right)
\left( 1-\delta _{m0}\right)\\
&~~~~ +\delta _{m0}p_{0}\left( 0\right) +\delta
_{m1}p_{1}\left( 0\right)   \label{RR1} \\
3p_{1}\left( m\right)  &=p_{0}\left( m\right) +p_{1}\left( m+1\right)\\
&~~~~+p_{1}\left( m-1\right) \left( 1-\delta _{m0}\right) +\delta
_{m0}p_{1}\left( 0\right)   \label{RR2}
\end{split}
\end{eqnarray}%
A simplification occurs when we sum the two sets (balancing the currents
between the $p_{0}$'s and the $p_{1}$'s): 
\begin{equation}
\sum_{m\geq 0}p_{0}\left( m\right) =\sum_{m\geq 0}p_{1}\left( m\right) =%
\frac{1}{2}  \label{sum}
\end{equation}%
the last ``='' being the result of normalization. To obtain the individual $p$'s, we
consider the generating functions:
\begin{equation}
G_{\bullet }\left( z\right) =\sum_{m\geq 0}z^{m}p_{\bullet }\left( m\right) 
\end{equation}%
and verify that they satisfy 
\[
\left( 
\begin{array}{cc}
1-2z & z \\ 
1 & 1-z%
\end{array}%
\right) \left( 
\begin{array}{c}
G_{0} \\ 
G_{1}%
\end{array}%
\right) =\left( 
\begin{array}{c}
\left( 1-z\right) \left[ zp_{1}\left( 0\right) +p_{0}\left( 0\right) \right] 
\\ 
p_{0}\left( 0\right) +\left( 1+z\right) p_{1}\left( 0\right) 
\end{array}%
\right), 
\]%
where the second line expresses the balance of total fluxes between $m$ and $m+1$.
Thus, we have, e.g.,  
\begin{equation}
G_{0}=\frac{\left( 1-z\right) ^{2}\left[ p_{0}\left( 0\right) +zp_{1}\left(
0\right) \right] -z\left[ p_{0}\left( 0\right) +\left( 1+z\right)
p_{1}\left( 0\right) \right] }{2\left( z-\hat{z}\right) \left( z-1/2\hat{z}%
\right) }  \label{G0}
\end{equation}%
where $\hat{z}=1-1/\sqrt{2}$.  Since $G_{0}$ cannot be singular at $\hat{z}<1$, the numerator must vanish there and leads to a relation between $%
p_{0}\left( 0\right) $ and $p_{1}\left( 0\right) $:%
\begin{equation}
\left( 1-\hat{z}\right) ^{2}\left[ p_{0}\left( 0\right) +\hat{z}p_{1}\left(
0\right) \right)]-\hat{z}\left[ p_{0}\left( 0\right) +\left( 1+\hat{z}%
\right) p_{1}\left( 0\right) \right] =0  \label{zetap0q0}
\end{equation}%

A second relation between them comes from Eq.~(\ref{sum}),
\begin{equation}
1/2=G_{0}\left( 1\right) =p_{0}\left( 0\right) +2p_{1}\left( 0\right) 
\label{sum1/2}
\end{equation}
and allows us to arrive at:

\begin{eqnarray}
p_{0}\left( 0\right) &=&\frac{11-6\sqrt{2}}{14}\approx 0.1796\\
p_{1}\left( 0\right) &=& \frac{3\sqrt{2}-2}{14}\approx 0.1602.
\end{eqnarray}

Instead of writing explicit expressions for all the $p$'s, let us exploit a
shortcut to the asymptotic behavior, namely, subtract (\ref{zetap0q0}) from
the numerator in (\ref{G0}) and cancel the $\left( z-\hat{z}\right) $ in the
denominator. The result is that $G_{0}\left( z\right) $ must be of the form%
\begin{equation}
G_{0}\left( z\right) =\frac{A+Bz+Cz^{2}}{1-2\hat{z}z}
\label{g0}
\end{equation}%
where $A,B,C$ are constants that can be explicitly computed.
From here we find that, for all $m\geq 2$,
\begin{equation}
p_{0}\left( m\right) =\zeta ^{m}\left\{ A+B/\zeta +C/\zeta ^{2}\right\} 
\end{equation}
where $\zeta \equiv 2\hat{z}=2-\sqrt{2}$. A similar expression can be derived for $p_1(m)$. Thus we see that the third hole is
bound with an exponential tail, of characteristic length $-1/\ln \zeta  \approx 1.8697$. 


The exact average of $m$, the distance between the engine and the caboose, can also be calculated via 
\begin{eqnarray}
\left<m\right> & = & \dfrac{d\left[G_0(z)+G_1(z)\right]}{dz}|_{z=1} \nonumber\\
 & = & \frac{8+9\sqrt{2}}{14}\approx 1.4806,
\label{avgm}
\end{eqnarray}
a result slightly smaller than the characteristic length of the exponential tail, $-1/\ln \zeta$.

As a cross check, we compute the average current explicitly, via 
\[
p_{0}\left( 0\right) +2p_{0}\left( 1\right) +2p_{0}\left( 2\right)
+...+3p_{1}\left( 0\right) +5p_{1}\left( 1\right) +5p_{1}\left( 2\right) +...
\]%
which is%
\[
2\sum_{m=0}p_{0}\left( m\right) +5\sum_{m=0}p_{1}\left( m\right)
-p_{0}\left( 0\right) -2p_{1}\left( 0\right) =3
\]%
by Eqs.~(\ref{sum}, \ref{sum1/2}). Since $H=3$, this result shows the UV property
explicitly.

Needless to say, if we reverse the order of limits ($\ell _{\max
}\rightarrow \infty $ followed by $L$ $\rightarrow \infty $), then the
system will be only in the AC phase, even though the exact $P^{\ast }$ is
yet to be obtained explicitly.

\bibliographystyle{apsrev4-1}
\bibliography{references_uv}

\end{document}